\documentclass[12pt, a4paper]{article}
\usepackage{epsf}
\usepackage{cite}
\usepackage{amsmath,amssymb}
\input{colordvi.tex}
\usepackage[usenames,dvipsnames]{color}
\usepackage{graphicx}
\usepackage{ascmac}
\usepackage{hyperref}
\usepackage{indentfirst}

\usepackage{bm}
\usepackage{braket}
\usepackage{cleveref}
\usepackage{siunitx}

\newcommand*{\tilda}[1]{{\tilde{a}_{\vec{#1}}}}
\newcommand*{\aket}[1]{{\ket{#1}_a}}

\begin{document}

\begin{titlepage}

\vspace{-4cm}
\begin{center}
\hfill TU-1206\\
\hfill KEK-QUP-2023-0026\\
\hfill KEK-TH-2561\\
\hfill KEK-Cosmo-0329\\
\hfill IPMU23-0037\\
\vskip 0.3in

{\Large \bf
Effects of Finite Material Size On Axion-magnon Conversion
}

\vskip .3in

{\large
So Chigusa$^{(a,b,c)}$,
Asuka Ito$^{(c,e)}$,
Kazunori Nakayama$^{(d,c)}$ \\
and
Volodymyr Takhistov$^{(c,e,f,g)}$
}

\vskip 0.3in

$^{(a)}${\em 
Berkeley Center for Theoretical Physics, Department of Physics, University of California, Berkeley, CA 94720, USA
}

\vskip 0.1in

$^{(b)}${\em 
Theoretical Physics Group, Lawrence Berkeley National Laboratory, Berkeley, CA 94720, USA
}

\vskip 0.1in

$^{(c)}${\em 
International Center for Quantum-field Measurement Systems for Studies of the Universe and Particles (QUP), KEK, 1-1 Oho, Tsukuba, Ibaraki 305-0801, Japan
}

\vskip 0.1in

$^{(d)}${\em 
Department of Physics, Tohoku University, Sendai 980-8578, Japan
}

\vskip 0.1in

$^{(e)}${\em 
Theory Center, Institute of Particle and Nuclear Studies (IPNS), High Energy Accelerator Research Organization (KEK), Tsukuba 305-0801, Japan
}

\vskip 0.1in

$^{(f)}${\em Graduate University for Advanced Studies (SOKENDAI), \\
1-1 Oho, Tsukuba, Ibaraki 305-0801, Japan}
 
\vskip 0.1in

$^{(g)}${\em 
Kavli Institute for the Physics and Mathematics of the Universe (WPI), UTIAS
The University of Tokyo, Kashiwa, Chiba 277-8583, Japan
} 

\end{center}
\vskip .15in

\begin{abstract}

Magnetic materials are particularly favorable targets for detecting axions interacting with electrons because the collective excitation of electron spins, the magnon, can be excited through the axion-magnon conversion process. It is often assumed that only the zero-momentum uniformly precessing magnetostatic (Kittel) mode of the magnon is excited. This is justified if the de Broglie wavelength of the axion is much longer than the size of the target magnetic material. However, if the de Broglie wavelength is shorter, finite-momentum magnon modes can also be excited. We systematically analyze the target material size dependence of the axion-magnon conversion rate. We discuss the importance of these effects in the detection of relativistic axions as well as in the detection of axion dark matter of relatively heavy mass with large material size.

\end{abstract}

\end{titlepage}

\tableofcontents

\section{Introduction}

The axion can resolve the strong CP problem of the Standard Model and explain the reason for the apparent charge-parity symmetry of quantum chromodynamics (QCD), while at the same time providing a significant fraction of the observed dark matter (DM) abundance that is predominant form of matter in the Universe~\cite{Peccei:1977hh,Weinberg:1977ma,Wilczek:1977pj,Preskill:1982cy,Abbott:1982af,Dine:1982ah,Ipser:1983mw}. 
Discovery of the Higgs boson demonstrated that scalar (spin-0) fields definitively exist~\cite{ATLAS:2012yve,CMS:2012qbp} and existence of pseudoscalar axions and axion-like particles is expected from more fundamental theory~(e.g.~\cite{Arvanitaki:2009fg}).

Theories of QCD axion are broadly classified into the Kim-Shifman-Vainshtein-Zakharov (KSVZ) 
models~\cite{Kim:1979if,Shifman:1979if} that introduce additional heavy vector-like quarks and the Dine-Fischler-Srednicki-Zhitnitsky (DFSZ) models~\cite{Dine:1981rt,Zhitnitsky:1980tq} with an extended Higgs sector.
In both KSVZ and DFSZ classes of models, axions can directly interact with photons.
Broad experimental program has been established to search for the axions coupled to photons.
One example is ADMX (Axion Dark Matter eXperiment), which employs a 
haloscope utilizing powerful magnetic fields to convert axions from Galactic DM halo into detectable microwave photons within a cavity~\cite{ADMX:2018gho,ADMX:2019uok,ADMX:2021nhd}. 
On the other hand, ABRACADABRA (A Broadband/Resonant Approach to Cosmic Axion Detection with an Amplifying B-field Ring Apparatus)~\cite{Salemi:2021gck} haloscope experiment represents a resonance search approach that enhances sensitivity to axion detection by utilizing a superconducting ring with a strong magnetic field.
In contrast, helioscopes
explore the possibility of axions originating from the Sun.
As an example, CAST (CERN Axion Solar Telescope)~\cite{CAST:2017uph} observes axion-to-photon conversion in the strong magnetic field of the telescope.

Aside from coupling to photons, axions in the DFSZ models directly couple to electrons at the tree level.
Such axion coupling to electrons also naturally appears within flavorful axion models~\cite{Ema:2016ops,Calibbi:2016hwq}, where axion physics is linked to flavor symmetries.
Therefore, probing the axion-electron coupling is also crucial for distinguishing QCD axion models.
One possible method for investigating the axion-electron interactions is detection of magnons, quasi-particles consisting of collective excitations of electron spins, in ferromagnetic crystals~\cite{BARBIERI1989357,Chigusa:2020gfs,Mitridate:2020kly}. 
Several experiments have already  conducted a search for excitation of magnons by axions with 
a linear amplifier~\cite{Crescini:2018qrz,Flower:2018qgb,QUAX:2020adt} as well as with
a quantum nondemolition detection technique~\cite{Ikeda:2021mlv}.

Previous works that considered magnons for axion DM detection have focused on excitation of the zero-momentum uniformly precessing magnetostatic (Kittel) magnon mode~\cite{Chigusa:2020gfs,Crescini:2018qrz,Flower:2018qgb,QUAX:2020adt,Mitridate:2020kly,Ikeda:2021mlv,Chigusa:2023hmz} (see also Refs.~\cite{Marsh:2018dlj,Schutte-Engel:2021bqm,Chigusa:2021mci} for similar ideas utilizing topological magnetic material).
This is a valid approach when the typical size of the target magnetic material is longer than the axion de Broglie wavelength $\lambda_a$, which is calculated for the axion DM as $\lambda_a = 2\pi (m_a v_a)^{-1} \simeq 1\,{\rm m}\,(m_a/1\,{\rm meV})^{-1}$ with $m_a$ being the axion mass. 
However, in various situations and parameter space corners where Kittel mode can be not the dominant excited mode. Two prominent scenarios include:
\begin{itemize}
\item The axions are much heavier with $m_a \gg 1\,{\rm meV}$ and its de Broglie wavelength can be shorter than the target material size. Hence, the axion detection sensitivity can be improved by taking into account the high-momentum magnon excitations in a non-resonant broadband search. 
\item The relativistic axions that can be produced by a variety of distinct processes, including cosmological ones. Such axions will excite high-momentum magnon modes rather than the Kittel mode.
\end{itemize}

In this paper, we present a formulation of the magnon excitation rate due to incoming axions with general momenta for general material size~\footnote{Refs.~\cite{Trickle:2019ovy,Esposito:2022bnu} considered a detection of MeV-scale scattering DM with magnons.}. Our formalism clearly illustrates how the notion of ``momentum conservation'' plays a role when the target material size is varied. Transition from the case where the Kittel mode is dominant and the case where the higher-momentum modes are important is explicit from this point of view.

This paper is organized as follows.
In Sec.~\ref{sec:dm}, we calculate the magnon excitation rate for the axion cold DM taking account of the finite material size effect and finite momentum magnon modes.
In Sec.~\ref{sec:rel}, we calculate the magnon excitation rate for axions with more general momentum distribution, including the axion dark radiation. For that purpose, we use the quantum mechanical treatment of axion. 
We conclude in Sec.~\ref{sec:con}.

\section{Conversion of dark matter axions} \label{sec:dm}

\subsection{Formulation}

Consider that the axion can interact with electrons (e.g. DFSZ~\cite{Dine:1981rt,Zhitnitsky:1980tq} or flavorful axion~\cite{Ema:2016ops,Calibbi:2016hwq} models), through the Lagrangian density coupling $\mathcal{L} \supset (\partial_{\mu} a/2 f)\overline{\psi_e}\gamma^{\mu}\gamma_5\psi_e$ where $\psi_e$ denotes the electron field and $f$ the scale of the Peccei-Quinn symmetry breaking. In the non-relativistic limit of $\psi_e$, the corresponding interaction Hamiltonian can be expressed with electron spin operators $\vec{S}_\ell$ as
\begin{align}
	H_{\rm int} = \frac{1}{f}\sum_{\ell} \vec\nabla a(\vec x_\ell)\cdot \vec S_\ell~.
	\label{Hint}
\end{align}
Here, we assume an magnetic material with electron spins localized at the lattice points labelled by $\ell$ are aligned.
The axion field can be treated classically as
\begin{align}
	a(t,\vec x)= a_0 \cos(m_at-m_a\vec v_a\cdot\vec x+\delta)~,
	\label{axion}
\end{align}
which holds within the axion coherence time $\tau_a \sim (m_a v_a^2)^{-1}$. For axion DM, $m_a a_0 = \sqrt{2 \rho_{\rm DM}}$ where $\rho_{\rm DM} = \SI{0.3}{GeV / cm^3}$ is the local DM density. 

The electron spin is expressed by using the magnon creation/annihilation operator $c_\ell$ and $c_\ell^\dagger$
through the Holstein-Primakoff transformation:
\begin{align}
	S_\ell^x \equiv \sqrt{\frac{s}{2}}(c_\ell + c_\ell^\dagger),~~~S_\ell^y \equiv -i\sqrt{\frac{s}{2}}(c_\ell - c_\ell^\dagger),~~~
	S_\ell^z \equiv s- c_\ell^\dagger c_\ell,
\end{align}
to the lowest order in $c_\ell$ and $c_\ell^\dagger$. The creation/annihilation operators satisfy the commutation relation
\begin{align}
	\left[ c_\ell, c_{\ell'}^\dagger \right] = \delta_{\ell\ell'}.
\end{align}
The interaction Hamiltonian (\ref{Hint}) is rewritten as
\begin{align}
	H_{\rm int} = \frac{a_0 m_a v_a^+}{2if}\sqrt{\frac{s}{2}}\sum_{\ell} c_\ell^\dagger\left[e^{-i\vec k_a\cdot \vec x_\ell}e^{i(m_at+\delta)} - {\rm h.c.}\right], 
\end{align}
where we have defined $v_a^+\equiv v^x_a+iv^y_a$ and we dropped a term proportional to $c_\ell$ because it vanishes when acting on the vacuum state $\left|{\rm vac}\right>$ and does not contribute to the axion-magnon conversion.

The discrete Fourier transformation of the magnon creation/annihilation operator is defined by\footnote{
The definition of sign of $\vec k$ is different from Ref.~\cite{Chigusa:2020gfs}.
}
\begin{align}
	c_\ell = \frac{1}{\sqrt N} \sum_{\vec k} e^{i\vec k\cdot \vec x_\ell} c_{\vec k},~~~~~~~
	c_\ell^\dagger = \frac{1}{\sqrt N} \sum_{\vec k} e^{-i\vec k\cdot \vec x_\ell} c^\dagger_{\vec k},
\end{align}
where $N$ is the total number of the lattice points in the material. They also satisfy
\begin{align}
	\left[ c_{\vec k}, c_{\vec k'}^\dagger \right] = \delta_{\vec k\,\vec k'}.
\end{align}
Using this, the interaction Hamiltonian is further rewritten as
\begin{align}
	H_{\rm int} = \frac{a_0 m_a v_a^+}{2if}\sqrt{\frac{sN}{2}}\sum_{\vec k} c_{\vec k}^\dagger\left[F(\vec {q'}) e^{i(m_at+\delta)} - F(\vec {q}) e^{-i(m_at+\delta)}\right], 
\end{align}
where $\vec {q'} \equiv -\vec k-\vec p$ and $\vec {q}\equiv -\vec k+\vec p$ with $\vec p$ being the axion momentum, and we defined the form factor
\begin{align}
	F(\vec q) \equiv \frac{1}{N}\sum_{\ell} e^{i\vec q\cdot \vec  x_\ell}.
	\label{F}
\end{align}
Note the (approximate) normalization condition $\sum_{\vec q} \left| F(\vec q) \right|^2 = 1$.
It is easily seen that it gives $F(0)=1$ for $\vec q=0$ and this should be the maximum since, for nonzero $\vec q$, there are cancellations in the summation. 
One can naively estimate that, for $|\vec q| \lesssim 1/L$, where $L$ is the typical size of the material, the summation still coherently adds up and hence $F(\vec q)\sim 1$ for $|\vec q|\lesssim 1/L$.
Below, we further investigate this function to see this more explicitly.

In order to see the behavior of $F(\vec q)$, let us take the continuum limit, $N\to\infty$, with fixing the size of the material. Then, Eq.~(\ref{F}) is approximated by the integral
\begin{align}
	F(\vec q) = \frac{1}{V} \int_V d^3 x\,e^{i\vec q\cdot \vec  x}.
\end{align}
It is evident that this gives a delta function in the $V\to\infty$ limit, which is nothing but a momentum conservation as expected from the translational invariance; it requires that the magnon momentum should be exactly equal to the incoming axion momentum, $\vec k=\vec k_a$. 
With finite volume, however, there is no exact translational invariance and the momentum conservation is violated. 
Note that there is an (approximate) normalization condition:
\begin{align}
	V\int\frac{d^3 q}{(2\pi)^3} \left| F(\vec q) \right|^2 = 1. 
\end{align}

\begin{figure}
\begin{center}
   \includegraphics[width=10cm]{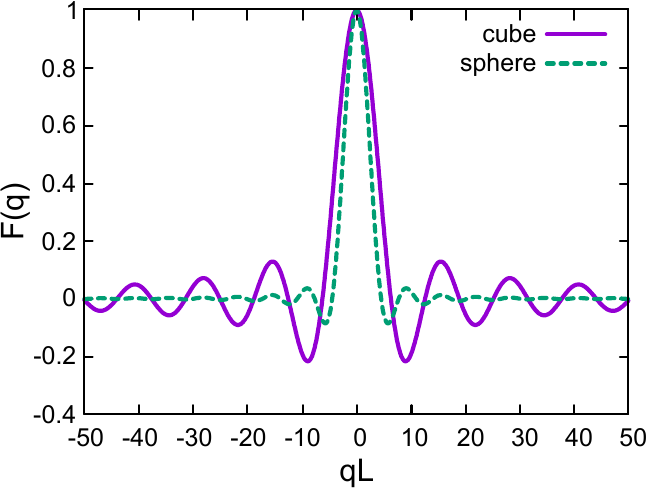}
  \end{center}
  \caption{The form factor $F(\vec q)$ as a function of $qL$ for cubic (purple) and spherical (green) materials. For the cubic material, we have taken $q_x=q$ and $q_y=q_z=0$. }
  \label{fig:Fq}
\end{figure}

For more concrete discussion, let us consider two representative shapes of the material: cube and sphere. 
For a cube with the side length $L$, $F(\vec{q})$ is explicitly calculated as
\begin{align}
	F(\vec q) = \frac{\sin(q_xL/2)}{q_xL/2} \frac{\sin(q_yL/2)}{q_yL/2} \frac{\sin(q_zL/2)}{q_zL/2}.
    \label{F_cubic}
\end{align}
For a sphere with radius $L$, it is calculated as\footnote{
    For a spherical magnetic material, the magnetic fluctuation should be expanded in terms of the so-called Walker (magnetostatic resonance) modes~\cite{PhysRev105390,fletcher1959ferrimagnetic}, instead of the Fourier modes, if the typical magnon momentum is comparable to the inverse of the material size.
}
\begin{align}
	F(\vec q) = \frac{3\left[\sin(qL)-qL \cos(qL)\right]}{(qL)^3}.
\end{align}
In both cases, it is peaked at $\vec q=0$, at which $F(0)=1$, but has a typical width of $\sim 2\pi/L$. The shape of $F(\vec{q})$ is shown in Fig.~\ref{fig:Fq}.
Roughly speaking, the momentum conservation is violated at the level of $\sim 2\pi/L$.
It should be compared with the typical separation of the discretized momenta inside the material, which is also of the order of $2\pi/L$. 
Therefore, if the axion momentum $k_a$ is smaller than $2\pi/L$, or the de Broglie wavelength is longer than $L$, the dominant excited magnon mode is the $\vec k=0$ mode, i.e., the Kittel mode. 
Since the de Broglie wavelength is given by $\lambda_a = 2\pi (m_a v_a)^{-1} \simeq 1\,{\rm m}\,(m_a/1\,{\rm meV})^{-1}$, we should take account of the magnon excitation with finite momentum for a future detector with a large material size, or for relatively heavy axion search.

\subsection{Conversion rate}  \label{sec:conv}

The total Hamiltonian for the axion and magnon is 
\begin{align}
	H = H_0 + H_{\rm int},
\end{align}
where $H_0$ is the magnon Hamiltonian, which comes from the Heisenberg interaction, given by
\begin{align}
	H_0 = \sum_{\vec k} \omega_k\, c_{\vec k}^\dagger c_{\vec k}.
\end{align}
Here, $\omega_k$ denotes the magnon dispersion relation;\footnote{
In general, the magnon frequency depends on the direction of the magnon momentum $\vec{k}$.
Below, we assume that $\omega_k$ only depends on the absolute value of the momentum $k$ just for simplicity of discussion.
} in particular, for the acoustic magnon mode with the lowest frequency, $\omega_0 = \gamma_e B_0$ is the Larmor frequency determined solely by the gyromagnetic ratio of electron $\gamma_e$ and the external magnetic field $B_0$. For the acoustic magnon mode with $k \ll 2\pi/L$, we obtain the quadratic momentum dependence $\omega_k \simeq \omega_0 + k^2/2M$ with $M \sim 3.5\,{\rm MeV}$ for the Yttrium Iron Garnet (YIG).

To calculate the conversion rate into the magnon mode with momentum $\vec k$, let us expand the magnon state as
\begin{align}
	\left|\psi(t)\right>=\alpha_{\rm vac} (t) \left|{\rm vac}\right> + \sum_{\vec k} \alpha_{\vec k} (t) \left|\vec k\right>,
\end{align}
with $ \left|\vec k\right> = c^\dagger_{\vec k}\left|{\rm vac}\right>$. We assume an initial condition $\alpha_{\rm vac} (0)=1$ and $\alpha_{\vec k} (0)=0$.\footnote{
	We express the vacuum state (or no magnon state) by $\left|{\rm vac}\right>$, in order to avoid the confusion with the state with one Kittel mode excitation: $\ket{\vec{0}}=c^\dagger_{\vec 0} \left|{\rm vac}\right>$.
}
By defining the interaction picture state $\left|\widetilde\psi(t)\right>\equiv e^{iH_0t} \left|\psi(t)\right>$, the Schrodinger equation is given by
\begin{align}
	i \frac{d}{dt}\left|\widetilde\psi(t)\right> = e^{iH_0t} H_{\rm int} e^{-iH_0t} \left|\widetilde\psi(t)\right>.
	\label{Schrodinger}
\end{align}
Under the approximation $\alpha_{\rm vac}\simeq 1$, it leads to
\begin{align}
	i\frac{d}{dt} \left( e^{i\omega_k t} \alpha_{\vec k}(t) \right) = 
	\frac{\sqrt{sN \rho_a} v_a^+}{2if}\left[F(\vec {q'}) e^{i(\omega_k +m_a)t}e^{i\delta} - F(\vec {q}) e^{i(\omega_k-m_a)t} e^{-i\delta}\right].
	\label{ddt_alphak}
\end{align}
Note that the second term gives the resonance and the dominant contribution for $\omega_k \simeq m_a$. Taking the $t\to \infty$ limit, this represents the energy conservation in the limit of vanishing dissipation.
In a real material, the energy conservation is only approximate because there is a finite magnon width.
We can analytically integrate this equation as
\begin{align}
	\alpha_k(t) = C \frac{F(\vec {q'})e^{i\delta}(m_a-\omega_k)(e^{im_at}-e^{-i\omega_kt}) + F(\vec {q})e^{-i\delta}(m_a+\omega_k)(e^{-im_at}-e^{-i\omega_kt})}{\omega_k^2-m_a^2},
\end{align}
where $C\equiv \frac{\sqrt{sN \rho_a} v_a^+}{2if}$. Around the resonance frequency $\omega_k \simeq m_a$, the first term oscillates with the time scale $m_a^{-1}$ while the second term grows linearly with time until $t \lesssim |m_a-\omega_k|^{-1}$.
To the leading order in $|m_a-\omega_k|$, we obtain
\begin{align}
	\alpha_k(t) \simeq -iCF(\vec {q})e^{-i(m_a t+\delta)} t.
\end{align}
Thus, the conversion probability to the one magnon state $\left|\vec k\right>$ per unit time is
\begin{align}
	P_{\vec k}(t) = \frac{\left|\alpha_{\vec k}(t)\right|^2}{t}
	\simeq |C|^2 \left| F(\vec {q})\right|^2 t
	= \frac{sN \rho_a |v_a^+|^2}{4f^2}\left| F(\vec {q})\right|^2 t.
	\label{P_dm}
\end{align}
This is the same expression as that given in Ref.~\cite{Chigusa:2020gfs} up to the form factor $\left| F(\vec {q})\right|^2$. The information of the finite material size effect is contained in this factor.

\begin{figure}
\begin{center}
   \includegraphics[width=12cm]{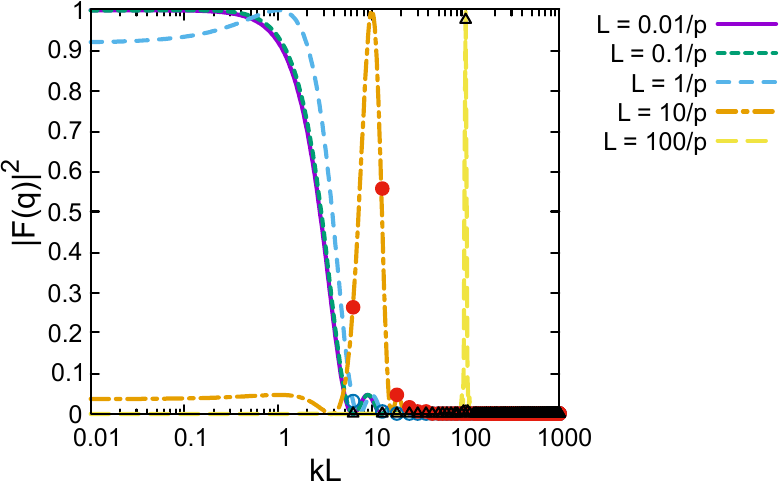}
  \end{center}
  \caption{The square of the form factor $|F(\vec q)|^2$, where $\vec q=\vec p-\vec k$ with $\vec p$ and $\vec k$ being the axion and magnon momenta, respectively, as a function of $kL$ for several choice of the material size: $Lp = 0.01, 0.1, 1, 10, 100$. For $Lp = 1, 10, 100$, points corresponding to discrete momenta with $kL=2\pi n$ $(n=1,2\dots)$ are also plotted. }
  \label{fig:Fq2}
\end{figure}

Fig.~\ref{fig:Fq2} shows $\left| F(\vec {q})\right|^2$ as a function of $kL$ with $k$ being the magnon momentum for several choices of the material size: $Lp = 0.01, 0.1, 1, 10$ with $p = m_a v_a$ being the axion momentum. We have assumed a cubic target material as described by Eq.~(\ref{F_cubic}) and considered the axion momentum parallel to the $x$-direction by taking $p_y=p_z=0$ for simplicity.
For $Lp = 1, 10, 100$, points at discrete momenta $kL=2\pi n$ $(n=1,2,\dots)$ are also plotted. 
Note that the $n=0$ points (the Kittel mode) are not shown in the figure since the horizontal axis is logarithmic scale.
It can be seen that, for a small material size $Lp \lesssim 1$, only magnons with a momentum $k\lesssim 2\pi/L$ are excited. Since the typical separation of the adjacent modes in the momentum space is $\sim 2\pi/L$, this means that only the Kittel mode $\vec k=0$ ($n=0$) is excited in this limit.
On the other hand, for a larger material size $L p \gtrsim 1$, larger momentum modes around $k\sim p$ are excited while the excitation of the modes $k < 1/L$ is suppressed.
For very high axion momentum or material size $L p \gg 1$, it can be seen that almost only one discrete magnon momentum mode with $k\simeq p$ is excited. 
The typical width of the peak is $\sim 2\pi/L$ around $k=p$, and hence the momentum conservation becomes more accurate for larger $L$. 

Let us comment on the detection probability taking into account the material size effect. This sensitively depends on the particular experimental setup, or how magnons are detected. If only the Kittel mode excitation is analyzed, one should derive $P_{\vec k=0}(t)$, which is proportional to the form factor $\left| F(\vec{q}=\vec{p})\right|^2$. As is clear from Fig.~\ref{fig:Fq2} or Eq.~(\ref{F_cubic}), it is highly suppressed as $\sim (pL)^{-6}$ for $p L \gg 1$. 
This applies to the case of Ref.~\cite{Ikeda:2021mlv}, for example, where a magnetic material is placed in a cavity and the Kittel mode of magnon is detected through the qubit system.
On the other hand, if magnon detection is performed inclusively, we should sum up all the magnon frequency modes. In such a case, due to the normalization condition $\sum_{\vec q} \left| F(\vec q) \right|^2 = 1$, one can simply replace the form factor with unity in Eq.~(\ref{P_dm}).
A possible example of the inclusive setup is the calorimetric magnon detection proposed in Ref.~\cite{Trickle:2019ovy}.

\section{Conversion of axions: quantum treatment} \label{sec:rel}

\subsection{Formulation}

In the previous section, we treated axion DM as a classical field.  
Here, let us consider a quantum treatment of the axion. The axion field is quantized as follows:
\begin{align}
  a(\vec{x},t) = \int \frac{d^3 \vec{p}}{(2\pi)^3} \frac{1}{\sqrt{2\omega_p}}
  \left(
    a (\vec{p}) e^{i (\vec{p}\cdot \vec{x}-\omega_p t)} + \mathrm{h.c.}
  \right),
\end{align}
with $\omega_p^2 \equiv p^2 + m_a^2$.
The operators satisfy the bosonic commutation relation
\begin{align}
  \left[
    a(\vec{k}), a(\vec{p})^\dagger
  \right] = (2\pi)^3 \delta^{(3)} (\vec{k} - \vec{p}).
\end{align}
Sometimes, it is also convenient to consider the quantization within a box with finite volume $\mathcal V$, which gives us
\begin{align}
  a(\vec{x},t) = \frac{1}{\mathcal V} \sum_{\vec{p}} \frac{1}{\sqrt{2\omega_p}}
  \left(
    a_{\vec{p}}\, e^{i(\vec{p}\cdot\vec{x}-\omega_pt)} + \mathrm{h.c.}
  \right),
\end{align}
with
\begin{align}
  \left[
    a_{\vec{k}}, a_{\vec{p}}^\dagger
  \right] = \mathcal V \delta_{\vec{k}, \vec{p}},
\end{align}
where the momenta $\vec{k}$ and $\vec{p}$ are chosen such that the corresponding wave function is consistent with the boundary conditions.
Note that we can take a continuum limit by replacing $\frac{1}{\mathcal V}\sum_{\vec{p}}$ with $\int \frac{d^3 \vec{p}}{(2\pi)^3}$ and $\mathcal V\delta_{\vec{k}, \vec{p}}$ with $(2\pi)^3 \delta^{(3)} (\vec{k} - \vec{p})$.
We also define dimensionless operators $\tilda{p} \equiv \mathcal V^{-1/2} a_{\vec{p}}$, with which we can rewrite the axion field as
\begin{align}
  a(\vec{x},t) = \frac{1}{\sqrt{\mathcal V}} \sum_{\vec{p}} \frac{1}{\sqrt{2\omega_p}}
  \left(
    \tilda{p} \, e^{i(\vec{p}\cdot\vec{x}-\omega_p t)} + \mathrm{h.c.}
  \right),
\end{align}
where
\begin{align}
  \left[
    \tilda{k}, \tilda{p}^\dagger
  \right] = \delta_{\vec{k}, \vec{p}}.
\end{align}
Using the dimensionless operators $\tilda{p}$, we define the axion Fock space.
In particular, we define the vacuum state $\aket{0}$ with $\tilda{p}\, \aket{0} = 0$ for any $\vec{p}$, while the one-axion state is defined as $\aket{\vec{p}} \equiv \tilda{p}^\dagger\, \aket{0}$.

Let us define the $\mathcal N_{\vec{p}}$-particle state of a fixed momentum $\vec{p}$ as
\begin{align}
	\left| \mathcal N_{\vec{p}} \right> \equiv \frac{e^{i\theta_{\vec p}}}{\sqrt{\mathcal N_{\vec{p}}!}} \left(\tilde a_{\vec{p}}^\dagger\right)^{\mathcal N_{\vec{p}}}  \left| 0 \right>.
\end{align}
It satisfies $\tilde a_{\vec{p}}^\dagger \left| \mathcal N_{\vec{p}} \right>  = \sqrt{\mathcal N_{\vec{p}}+1} \left| \mathcal N_{\vec{p}}+1\right>$ and $\tilde a_{\vec{p}} \left| \mathcal N_{\vec{p}} \right>  = \sqrt{\mathcal N_{\vec{p}}} \left| \mathcal N_{\vec{p}}-1\right>$.
The probability amplitude of going from the $\left| \mathcal N_{\vec{p}} \right> $ state to the $\left| \mathcal N_{\vec{p}}-1 \right> $ state is given by
\begin{align}
	\left< \mathcal N_{\vec{p}}-1\right| a(t,\vec x)\left| \mathcal N_{\vec{p}} \right> =  \sqrt{\frac{\mathcal N_{\vec{p}}}{2\omega_p \mathcal V}} e^{-i(\omega_pt - \vec p\cdot\vec x-\theta_{\vec p})}.
\end{align}

\subsection{Conversion rate} \label{sec:conv_rel}

The axion-magnon interaction Hamiltonian of Eq.~(\ref{Hint}) is expressed as
\begin{align}
	H_{\rm int}=\sum_{\vec p} \sum_{\vec k} \frac{1}{f}\sqrt{\frac{sN}{4\omega_p\mathcal V}} i p^+ c_{\vec k}^\dagger
	\left[  \tilde a_{\vec{p}} F(\vec q) e^{-i\omega_p t} -  \tilde a_{\vec{p}}^\dagger F(\vec {q'}) e^{i\omega_p t}\right],
\end{align}
where $p^+\equiv p_x+ip_y$, $\vec q \equiv -\vec k + \vec p$, and $\vec{q'}\equiv -\vec k-\vec p$ with $\vec k$ being the magnon momentum.
Here, we only picked up terms proportional to $c_{\vec k}^\dagger$ since other terms do not contribute to the axion-magnon conversion.
Also, note that the summations of $\vec{p}$ and $\vec{k}$ run over different sets of momenta since the axion and magnon are quantized in different boxes with volume $\mathcal{V}$ and $V$, respectively.
Suppose that the initial state is given as $\left|\mathcal N_{\vec{p}}; {\rm vac}\right>$ where $\mathcal N_{\vec{p}}$ represents the axion $\mathcal N_{\vec{p}}$-particle state and ``vac'' represents the ground state of the magnetic material without magnons.
At the level of the first order perturbation in $H_{\rm int}$, one magnon is excited and one axion is created or annihilated. However, we can safely neglect the axion creation since the time integral of the Schrodinger equation gives almost zero due to the rapid oscillation as in the case of Sec.~\ref{sec:conv}.
This is consistent with the expected energy conservation.
Therefore, we can just focus on the term proportional to the annihilation operator $\tilde a_{\vec{p}}$.

Therefore, the general state of our interest is expressed as
\begin{align}
	\left| \psi(t) \right> = \alpha_{\vec p,{\rm vac}}(t)\left|\mathcal N_{\vec{p}}; {\rm vac}\right> + \sum_{\vec k}\alpha_{\vec p, \vec k}(t) \left|\mathcal N_{\vec{p}}-1; \vec k\right>.
 \label{psi_t}
\end{align}
By using the Schrodinger equation (\ref{Schrodinger}), we obtain 
\begin{align}
	i\frac{d}{dt}\left( e^{i\omega_k t} \alpha_{\vec p,\vec k}(t) \right)
	=\frac{ i \alpha_{\vec p,{\rm vac}}(t) \, p^+ }{f}\sqrt{\frac{sN\mathcal N_{\vec{p}}}{4\omega_p\mathcal V}} F(\vec q) e^{i(\omega_k-\omega_p)t}.
\end{align}
Note that this reduces to Eq.~(\ref{ddt_alphak}) if one takes the non-relativistic axion limit $p^+ = m_a v_a^+$, $\omega_p = m_a$ so that the energy density of axions with momentum $\vec{p}$ is given by $\rho_a(\vec{p}) = m_a \mathcal N_{\vec{p}} /\mathcal V$.\footnote{
    The coherently oscillating classical field may also be approximated by the coherent state~\cite{Matsumoto:2007rd}. The same result is obtained by using the coherent state, rather than the $\mathcal N$-particle state. 
}
For more general cases, assuming that $ \alpha_{\vec p,{\rm vac}}(t) \simeq 1$, we obtain
\begin{align}
	\left| \alpha_{\vec p,\vec k}(t)\right|^2 =\frac{s N \mathcal N_{\vec{p}} |p^+|^2 |F(\vec q)|^2}{f^2\omega_p \mathcal V}
	\frac{\sin^2\left((\omega_k-\omega_p)t/2\right)}{(\omega_k-\omega_p)^2}.
	\label{alpha_pk2}
\end{align}
For later convenience, we define $\alpha_{\vec p,\vec k}(t) = \sqrt{\mathcal{N}_{\vec{p}}}\, \tilde{\alpha}_{\vec p,\vec k}(t)$ so that $\tilde{\alpha}_{\vec p,\vec k}(t)$ does not depend on $\mathcal{N}_{\vec{p}}$.
Note that the factor of $\sin^2\left((\omega_k-\omega_p)t/2\right)/(\omega_k-\omega_p)^2$ in Eq.~(\ref{alpha_pk2}) is strongly peaked around $\omega_p=\omega_k$, whose typical width is given by $\sim 1/t$ for finite time $t$.

Now let us take into account the momentum distribution of the axion.
For each fixed momentum $\vec{p}$, the occupation number $\mathcal{N}_{\vec{p}}$ is distributed with some function $g_{\vec{p}}(\mathcal{N}_{\vec{p}})$, which is normalized as $\sum_{\mathcal{N}_{\vec{p}}}g_{\vec{p}} (\mathcal{N}_{\vec{p}}) = 1$.\footnote{
The axion state is a classical ensemble of the pure states $\ket{\{\mathcal{N}_{\vec{p}}\}}$ where $\{\mathcal N_{\vec p}\}$ represents a set of integers, with the distribution function $\prod_{\vec p}g_{\vec{p}} (\mathcal{N}_{\vec{p}})$ and conveniently expressed in the form of a density matrix.
See Appendix \ref{sec:dens} for details.
}
The averaged occupation number can be defined as 
\begin{align}
  \overline{\mathcal{N}}_{\vec{p}} \equiv \sum_{\mathcal{N}_{\vec{p}}} \mathcal{N}_{\vec{p}}\, g_{\vec{p}} (\mathcal{N}_{\vec{p}})
  \;.
\end{align}
Using the set of functions $\left\{ g_{\vec{p}}(\mathcal{N}_{\vec{p}}) \right\}$, the axion momentum distribution is expressed as
\begin{align}
  f(\vec{p}) = \frac{\overline{\mathcal{N}}_{\vec{p}}}{\overline{\mathcal{N}}}
  \;,
\end{align}
where $\overline{\mathcal{N}} \equiv \sum_{\vec{p}} \overline{\mathcal{N}}_{\vec{p}}$ is the averaged total number of axions.
$f(\vec{p})$ is normalized as $\sum_{\vec{p}} f(\vec{p}) = \mathcal{V} \int \frac{d^3p}{(2\pi)^3}\, f(\vec{p}) = 1$.
Then, the averaged probability to find the magnon state with $\vec k$ per unit time is given by
\begin{align}
	P_{\vec k}(t) &= \frac{1}{t} \sum_{\vec p} \sum_{\mathcal{N}_{\vec{p}}}
  g_{\vec{p}}(\mathcal{N}_{\vec{p}})\, \mathcal{N}_{\vec{p}}
  \left|\tilde{\alpha}_{\vec p,\vec k}(t)\right|^2
   \notag \\
  &= \frac{\overline{\mathcal{N}}}{t} \sum_{\vec p} f(\vec{p}) \left|\tilde{\alpha}_{\vec p,\vec k}(t)\right|^2
  \;.
	\label{P_kt_int}
\end{align}

For a non-relativistic axion, such as the DM axion, the distribution function satisfies $f(\vec{p}) \simeq 0$ for $p\gtrsim m_a$ and the summation over $\vec{p}$ can be limited to the values with $p \ll m_a$.
In this case, the axion frequency has a peaked distribution around $\omega_p \simeq m_a$.
If the peak width is narrower than $1/t$, we can expand the sine factor in Eq.~(\ref{alpha_pk2}) as
\begin{align}
  \frac{\sin^2\left((\omega_k-\omega_p)t/2\right)}{(\omega_k-\omega_p)^2} \simeq \frac{t^2}{4}
  \;,
\end{align}
so that $P_{\vec{k}} \propto t$ for $\omega_{k} \simeq \omega_p$.
This is consistent with the result in Sec.~\ref{sec:conv}. 
For a relativistic axion, we can use $\omega_p \simeq p$.
Assuming that the distribution function can be decomposed as $ f(\vec{p}) = \eta(\omega_p) \zeta(\Omega)$ with $\Omega$ the solid angle of the momentum $\vec{p}$ and that the typical width of $\eta(\omega_p)$ is much larger than $1/t$, we can safely take the $\eta(\omega_p)$ outside the summation in Eq.~(\ref{P_kt_int}) and just replace it with $\eta(\omega_k)$. 
Applying the same trick to the smooth function $|p^{+}|^2$ and the form factor $F(\vec{q})$, which requires $t/L \gg 1$, we obtain
\begin{align}
  P_{\vec k}(t) \simeq 
  \frac{s N}{4\pi f^2} \omega_k^3\, \overline{\mathcal{N}} \eta(\omega_k)
  \int \frac{d\Omega}{4\pi}\, \zeta(\Omega) |F(\vec q)|^2 \sin^2 \theta
  \;,
  \label{eq:relativi}
\end{align}
with $\vec{q} = (\omega_k \sin\theta \cos\phi, \omega_k \sin\theta \sin\phi, \omega_k \cos\theta)^T - \vec{k}$, where we used the identity
\begin{align}
  \int d\omega_p\, \frac{\sin^2\left((\omega_k-\omega_p)t/2\right)}{(\omega_k-\omega_p)^2} \simeq \frac{\pi t}{2}
  \;.  
\end{align}
This is nothing but the Fermi's golden rule.

Noting that the combination $\overline{\mathcal{N}} f(\vec{p})$ gives the phase space density of the axion, we can apply Eq.~\eqref{eq:relativi} to different sources of axions. As an example, for the case of cosmic thermal axion relic distribution (see Appendix~\ref{appssec:cosmicaxion}), we consider
\begin{align}
  \overline{\mathcal{N}} \eta(\omega_p) = \frac{1}{e^{\omega_p/T}-1}
  ~~;~~
  \zeta(\Omega) = 1
  \;.
\end{align}

\begin{figure}[tbp]
  \centering
  \includegraphics[width=\hsize]{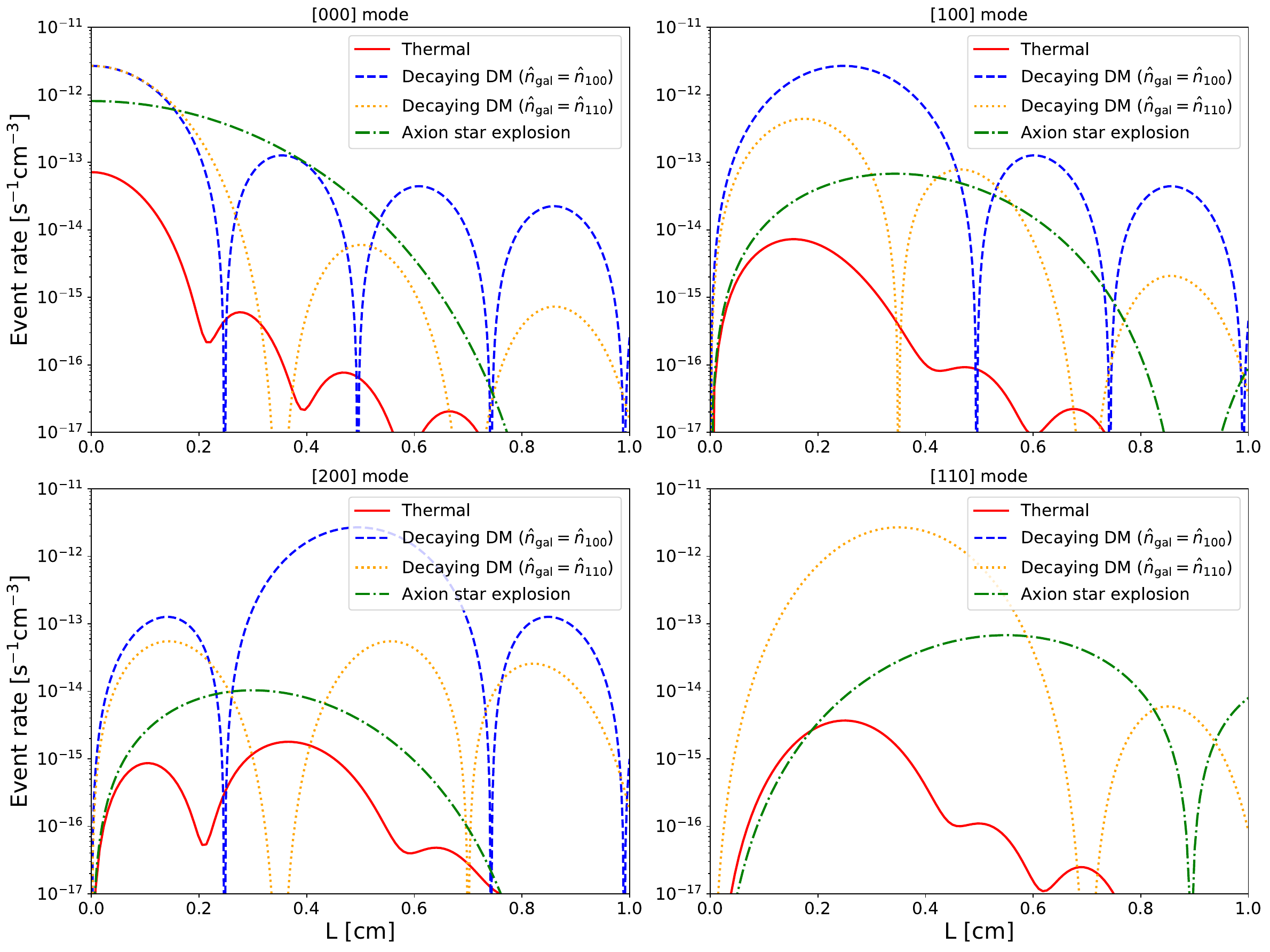}
  \caption{
    The magnon excitation rate per unit material volume as a function of the material size.
    Each panel corresponds to the Kittel mode (upper left), $[100]$ mode (upper right), $[200]$ mode (lower left), and $[110]$ mode (lower right).
    Lines correspond to axions sourced from the early Universe thermal bath (red solid), decaying Galactic DM with mass $m_\chi = \SI{1}{meV}$ under the material directions $\hat{n}_{\rm gal} = \hat{n}_{100}$ (blue dashed) and $\hat{n}_{\rm gal} = \hat{n}_{110}$ (orange dotted), and axion star explosion with $m_a=\SI{0.1}{meV}$ (green dash-dotted), respectively.
    The axion decay constant of $f_a = \SI{e10}{GeV}$ is assumed, while the external magnetic field $B_0$ is tuned for each setup such that the Larmor frequency matches the peak position of the axion spectrum.
  }
  \label{fig:comparison}
\end{figure}

In Fig.~\ref{fig:comparison}, we show the magnon excitation rate per unit material volume as a function of the material size $L$, where the dimensions of a cubic target material $L\times L\times L$ is assumed.
Each panel corresponds to the Kittel or $[000]$ mode (upper left), $[100]$ mode (upper right), $[200]$ mode (lower left), and $[110]$ mode (lower right) of the acoustic magnon in YIG, where the $[nm\ell]$ mode has the momentum $\vec{k}_{nm\ell}(L) = \frac{2\pi}{L} (n, m, \ell)^T$.
The red solid line denotes the contribution of thermal axions with temperature $T_a = T_0$, where the cosmic microwave background (CMB) temperature is given by $T_0 \simeq \SI{2.7}{K}$.
The blue dashed and orange dotted lines represent contributions of the Galactic component of decaying DM with mass $m_\chi = \SI{1}{meV}$, which is described in App.~\ref{sec:decayingDM}.
The difference between two setups is the direction of the Galactic Center $\hat{n}_{\rm gal}$ in terms of the coordinate system defined with the material axes, we use $\hat{n}_{\rm gal} = \hat{n}_{100}$ (blue dashed) and $\hat{n}_{\rm gal} = \hat{n}_{110}$ (orange dotted), where $\hat{n}_{nm\ell}$ is the unit vector pointing to the same direction with $\vec{k}_{nm\ell} (L)$.
The green dash-dotted lines show the contribution of axion star explosion located at $\mathcal{R}=\SI{1}{AU}$ from the Earth in the $\hat{n}_{111}$ direction with a fixed axion mass $m_a = \SI{1}{meV}$, which is described in App.~\ref{sec:axionStar}.
The axion decay constant of $f_a = \SI{e10}{GeV}$ is assumed for all setups.
Furthermore, the external magnetic field $B_0$ is tuned for each setup such that the corresponding Larmor frequency matches the peak position $\omega_0$ of the axion spectrum.
The concrete setups are as follows: $\omega_0 = \SI{0.65}{meV}$ and $B_0 = \SI{5.4}{T}$ for thermal axions, $\omega_0 = \SI{0.5}{meV}$ and $B_0 = \SI{4.2}{T}$ for axions from the decaying DM, and $\omega_0 = \SI{0.24}{meV}$ and $B_0 = \SI{2.2}{T}$ for axion star explosion.

The distinct shapes of event rate curves displayed in Fig.~\ref{fig:comparison} clearly demonstrate the non-trivial effects of material size, where widths of peak structures are inversely proportional to incoming axion momenta.
The early Universe thermal bath can be regarded as an example of the isotropic source of relativistic axions, while the emission from the Galactic component of the decaying DM and axion star explosion can be regarded as examples of directional sources.
From comparison between these two cases, it can be seen that the material size effects are more significant for directional sources, where the event rates can be suppressed by orders of magnitude when the material size corresponds to one of the ``dips'' of event rate curves.
On the other hand, the highest peaks of the curve for each magnon mode correspond to the positions where the two momenta, $\vec{p}$ and $\vec{k}_{nm\ell} (L)$, are close to each other.
For a fixed typical value of the axion momentum $p$, larger material size tends to increase the importance of magnon modes with larger $(n,m,\ell)$, as expected.

Comparison between blue dashed and orange dotted lines in Fig.~\ref{fig:comparison} provides us with intuition about approximate momentum conservation.
First, coincidence between two lines at the $L\to 0$ limit is expected since only the Kittel mode is excited in this limit, which is insensitive to the direction of the incoming momentum.
Secondly, the momentum conservation forces some of the setups to yield exactly zero rates.
Among these is the $[110]$ mode excitation from the decaying DM in the $\hat{n}_{100}$ direction. In this case, the incoming momentum does not have the $y$-component, which forbids excitation of any magnon modes with non-zero momentum along the $y$-axis.

\section{Conclusions and discussion} \label{sec:con}

The potential unique capabilities of magnetic materials for detecting axion-like particles or other new particles is drawing significant interest, and could prove to be essential for discriminating between different theoretical models after discovery.
We have systematically formulated the method for calculating magnon excitation rate taking into into account the effects of finite target material size.
As our formulation explicitly demonstrates, the axion excites only the lowest magnon material mode (Kittel mode) when the de Broglie wavelength is much longer than the target material size.
On the other hand, if the axion de Broglie wavelength is shorter than the target material size, high momentum magnons are readily excited. 
This behavior can be naturally understood as a consequence of the momentum conservation, since there is a translation invariance in the large target material size limit, while it is significantly broken in the small material size limit. 
These effects can be conveniently represented by the form factor $F(\vec q)$ (\ref{F}), whose precise functional form depends on the target material shape.

Our formulation is important from several viewpoints. 
Firstly, since the de Broglie wavelength of axion DM is given $\lambda_a \simeq 1\,{\rm m}\,(m_a/1\,{\rm meV})^{-1}$, it can be comparable or smaller than the target material size for relatively heavy axion in the future experiments with large magnetic materials. 
Secondly, magnetic materials may not be only sensitive to axion DM but also to relativistic axions that can appear from variety of distinct sources and our formulation provides a systematic approach to calculate the magnon excitation rate even for these cases.
Thirdly, the effects of finite material size we discuss could in principle be observed for other types of excitation modes in condensed matter systems, including phonons, if its detection method focuses on a specific momentum.
Finally, from an experimental side, our discussion serves as impetus and motivates further advances in developing optimal detailed methods for detection of high momentum magnon modes. Our formulation establishes a basis for estimating the possible detectable signals.

\section*{Acknowledgment}

This work was supported by the Director, Office of Science, Office of High Energy Physics of the U.S. Department of Energy under the Contract No. DE-AC02-05CH1123 (S.C.)
This work was supported by JSPS KAKENHI Grant Nos. 22K14034 (A.I.), 23K13109 (V.T.).
This work was supported by World Premier International Research Center Initiative (WPI), MEXT, Japan.

\appendix
\section{Sources of relativistic axions} \label{sec:app}

\subsection{Thermal cosmic axion background}
\label{appssec:cosmicaxion}

Relic axion abundance can be thermally produced in the early Universe and contribute to cosmic axion background~\cite{Turner:1986tb,Chang:1993gm,Masso:2002np,Hannestad:2005df,Graf:2010tv,Salvio:2013iaa,Arias-Aragon:2020shv,Dror:2021nyr}. Reminiscent of CMB, for some model interactions where axions are in thermal equilibrium with Standard Model constituents in the early Universe, the blackbody spectrum arises:
\begin{equation}
    \dfrac{d \rho}{d \log \omega} = \dfrac{1}{2 \pi^2} \dfrac{\omega^4}{e^{\omega/T_a}-1}~,
\end{equation}
where $T_a$ is the axion temperature at present time. Assuming that the spectrum originates from interactions that decoupled in the early Universe at temperature $T_d$ and that the entropy is conserved, one has
\begin{equation}
    T_a (T_d) \simeq T_0 \left(\dfrac{g_{\ast, S}(T_0)}{g_{\ast, S}(T_d)}\right)^{1/3}~,
\end{equation}
where $T_0 \simeq 2.7$~K is the present day temperature of the CMB and $g_{\ast, S}$ denotes the effective degrees of freedom associated with entropy.

We note that apart from thermal production, variety of mechanisms can lead to non-thermal cosmological axion production. 
For example, let us consider a complex scalar whose angular component corresponds to the (pseudo) Goldstone boson, axion. Its radial component is sometimes called saxion and it generally decays into the axion pair~\cite{Chun:1995hc,Asaka:1998xa,Ichikawa:2007jv,Kawasaki:2007mk}.
Axions can also be much more efficiently produced from non-linear dynamics of field oscillations compared to perturbative decays~\cite{Kasuya:1996ns,Ema:2017krp,Co:2017mop}.

\subsection{Decaying dark matter} \label{sec:decayingDM}

The nonthermal axions can be produced by decaying particles. An example is a scalar particle $\chi$ decaying into the axion pair as $\chi \to aa$, as mentioned in the previous subsection. 
In supersymmetric theories, gravitino (axino) can decay into the axino (gravitino) plus axion, where gravitino (axino) is the fermionic superpartner of the graviton (axion)~\cite{Chun:1993vz,Kim:1994ub,Asaka:2000ew}.
Here, we just assume that a scalar $\chi$ is stable on the cosmological time scale and constitutes the dominant component of DM and its small fraction is decaying into axion pairs~\cite{Dror:2021nyr}.

In this setup, the axion spectrum consists of the extragalactic and the Galactic components depending on where the scalar $\chi$ decays.
The logarithmic energy spectrum of the axion produced by the decaying DM, in terms of the density parameter normalized by the critical energy density of the present universe $\rho_c$, is given by
\begin{align}
  \Omega_a^{\rm ex} (\omega) &\simeq \Omega_\chi 
    \frac{2\omega}{m_\chi} \frac{1}{\tau H_0 \sqrt{\Omega_\Lambda + \Omega_m(1+z_d)^3}} \exp \left(
    - \frac{t(z_d)}{\tau} 
  \right) \Theta \left(
    \frac{m_\chi}{2} - \omega
  \right)\, \\
  \Omega_a^{\rm gal} (\omega) &\simeq \frac{\omega^2 e^{-t_U/\tau}}{2\pi m_\chi \tau \rho_c}
  D_{\mathrm{MW}} \delta \left(
    \frac{m_\chi}{2} - \omega
  \right),
  \label{Omegaa_DDM}
\end{align}
for the extragalactic and Galactic components, respectively, where $z_d = m_\chi/(2\omega)-1$ denotes the redshift at which the axion with the present energy $\omega$ was produced,\footnote{We assumed $z_d < z_{\rm eq}$, where $z_{\rm eq}$ is the redshift at the matter-radiation equality. It is valid unless we are interested in the very low energy tail of the axion spectrum.} $H_0$ is the present Hubble parameter, $\Omega_\Lambda \sim 0.7$ and $\Omega_m\sim 0.3$ are the density parameter of the cosmological constant and matter respectively~\cite{Planck:2018vyg}, $\tau$ is the lifetime of $\chi$, $t_U$ is the present age of the universe, $\rho_c$ is the critical energy density, and
\begin{align}
    t(z_d) = \frac{1}{3H_0 \sqrt\Omega_\Lambda}\log\left(\frac{\sqrt{\Omega_\Lambda + \Omega_m(1+z_d)^3}+\sqrt{\Omega_\Lambda}}{\sqrt{\Omega_\Lambda + \Omega_m(1+z_d)^3}-\sqrt{\Omega_\Lambda}}\right).
\end{align}
The first (second) term in (\ref{Omegaa_DDM}) denotes the extragalactic (Galactic) contribution. The extragalactic component has a broad energy spectrum peaked at $\omega = m_\chi/2$, while the Galactic one exhibits a delta function spectrum at $\omega=m_\chi/2$.

The axion phase space density can be evaluated using $\Omega_a(\omega)$ as
\begin{align}
  n(\vec{p}) = \frac{1}{4\pi} \frac{\rho_c}{\omega_a^4} \left(
    \Omega_a^{\rm ex} (\omega_a) + 4\pi \delta(\hat{p} + \hat{n}_{\rm gal}) \Omega_a^{\rm gal} (\omega_a)
  \right),
\end{align}
with $\omega_a \equiv |\vec{p}|$, $\hat{p} \equiv \vec{p}/|\vec{p}|$, and $\hat{n}_{\rm gal}$ being the unit vector towards the Galactic Center.

\subsection{Axion star explosions} \label{sec:axionStar}

Axion (boson) stars are solitonic gravitationally-bound
compact macroscopic objects, composed of ultralight axions (bosons)~\cite{Kaup:1968zz,Ruffini:1969qy,Colpi:1986ye,Seidel:1993zk}. In the early Universe, axion stars can form in cores of diffuse axion miniclusters~\cite{Kolb:1993zz,Eggemeier:2019jsu}.
Once they reach critical mass, their gravitational collapse results in exploding emission of significant amount of relativistic axions in the presence of sizable axion field self-interactions~\cite{Levkov:2016rkk,Eby:2016cnq,Helfer:2016ljl}. Relativistic axions from axion star explosions can lead to intriguing signatures in experiments searching for axion DM\footnote{Boson star explosions lead to novel signatures in quantum sensors~\cite{Arakawa:2023gyq}.}~\cite{Eby:2021ece}.

Let $\mathcal{E}$ be the total energy output of axion burst from a transient astrophysical source, such as axion star explosion.
The energy density at the detector $\rho_{*}$ can be evaluated as~\cite{Eby:2021ece}
\begin{align}
  \rho_{*} \simeq \frac{\mathcal{E}}{4\pi \mathcal{R}^2 \delta t},
\end{align}
where $\mathcal{R}$ is the distance from the detector to the axion star and $\delta t$ the observed burst duration.
Similarly, given the spectrum of emitted axions $d\mathcal{E}/dp$, the axion phase space density can be calculated as
\begin{align}
  n_a(\vec{p}) = \frac{1}{4\pi \mathcal{R}^2 \delta t}
  \frac{1}{p^2 \sqrt{p^2 + m_a^2}} \frac{d\mathcal{E}}{d p}
  \delta \left(
    \hat{p} + \vec{n}_{\mathrm{burst}}
  \right),
\end{align}
where $\hat{p}$ and $\vec{n}_{\mathrm{burst}}$ are the unit vectors along the axion momentum $\vec{p}$ and the line of sight to the axion star, respectively.

The properties of explosions, including the spectrum of emitted axions, has been obtained in simulations~\cite{Levkov:2016rkk}.
These simulations show an approximate relation $\delta t_{\rm burst} \simeq 400/m_a$ of the burst duration at the source.
Assuming that the wave spreading effect is negligible, we can use it to estimate $\delta t \simeq \delta t_{\rm burst}$.
Also, in this paper, we use a reference value from simulation results
\begin{align}
  \frac{d\mathcal{E}}{d p} \frac{m_a^2}{f_a^2 \mathcal{N}} \sim 10^3~,
\end{align}
which is typical for a mildly relativistic component with $p/m_a \sim O(1)$ including the peak position $p_0 \simeq 2.4 m_a$.
For estimation of the signal rate, we assume the number of explosions resulting in axion emission is $\mathcal{N}=1$.

\section{Density matrix formalism} \label{sec:dens}

In this Appendix, we use the density matrix formalism to reproduce the magnon excitation rate given in Sec.~\ref{sec:conv_rel} for general momentum distribution of the axion. The density matrix at the initial time $t=0$ is given by
\begin{align}
	\hat \rho (t=0) = \sum_{\{\mathcal N_{\vec p}\}} \Big( \prod_{\vec p} g_{\vec p}(\mathcal N_{\vec p}) \Big) \left| \{\mathcal N_{\vec p}\}; {\rm vac} \right>\left< \{\mathcal N_{\vec p}\}; {\rm vac} \right|,
\end{align}
where $\{\mathcal N_{\vec p}\} = (\mathcal N_1, \mathcal N_2 \dots)$ represents a set of integers where $\mathcal N_i$ is an abbreviation of $\mathcal{N}_{\vec{p}_i}$ that represents the number of particles with momentum $\vec p_i$.
The function $g_{\vec p}(\mathcal N_{\vec p})$ satisfies $\sum_{\mathcal N_{\vec p}} g_{\vec p}(\mathcal N_{\vec p})=1$. Note that ${\rm Tr}(\hat \rho)=1$.
To derive the magnon excitation rate, we calculate the expectation value of the magnon number operator $c_{\vec k}^\dagger c_{\vec k}$ with $\vec k$ being the magnon momentum.
To do so, we need to know the time evolution of the density matrix.
The density matrix evolves according to the Schrodinger equation (\ref{psi_t}):
\begin{align}
	 \left| \{\mathcal N_{\vec p}\}; {\rm vac} \right> \to \alpha_{\vec p,{\rm vac}}(t)\left| \{\mathcal N_{\vec p}\}; {\rm vac} \right> + \sum_{\vec p}\alpha_{\vec p,\vec k}(t)\left| \{\mathcal N_{\vec p}\}^{*}; \vec k \right>,
\end{align}
where $\alpha_{\vec p,\vec k}(t)$ is given in Eq.~(\ref{alpha_pk2}) and $\{\mathcal N_{\vec p}\}^* \equiv (\mathcal N_1,\mathcal N_2,\dots,\mathcal N_{\vec p}-1,\dots)$.
The latter term contributes to the magnon expectation value.
Thus, we obtain
\begin{align}
	{\rm Tr}\left(\hat\rho(t) \,c_{\vec k}^\dagger c_{\vec k}\right) &= \sum_{\{\mathcal N'_{\vec p} \},\lambda} \left< \{\mathcal N'_{\vec p}\}; \lambda\right|
	 \sum_{\{\mathcal N_{\vec p}\}}\prod_{\vec p} g_{\vec p}(\mathcal N_{\vec p})\left| \{\mathcal N_{\vec p}\}^*; \vec k \right> |\alpha_{\vec p,\vec k}(t)|^2 \left< \{\mathcal N_{\vec p}\}^*; \vec k\right| c_{\vec k}^\dagger c_{\vec k} \left| \{\mathcal N'_{\vec p}\}; \lambda \right> \nonumber\\
	 &= \sum_{\{\mathcal N_{\vec p} \}} \Big( \prod_{\vec p} g_{\vec p}(\mathcal N_{\vec p}) \Big)
	 \sum_{\vec p} \mathcal N_{\vec p} |\tilde\alpha_{\vec p,\vec k}(t)|^2 \nonumber\\
	 &=\sum_{\vec p}\overline{\mathcal N}_{\vec p}  |\tilde\alpha_{\vec p,\vec k}(t)|^2.
    \label{Tr_rho}
\end{align}
In the first line, the trace over the magnon state is assumed: $\lambda={\rm vac}, \vec k$.
We have defined the momentum distribution, or the averaged occupation number, as
\begin{align}
	\overline {\mathcal N}_{\vec p} = \sum_{\mathcal N_{\vec p}} \mathcal N_{\vec p}\,g_{\vec p}(\mathcal N_{\vec p}).
\end{align}
By dividing the result (\ref{Tr_rho}) with time $t$, we reproduce the result (\ref{P_kt_int}).

For the case of thermal distribution, we have $g_{\vec p}(\mathcal N_{\vec p}) = e^{-\omega_p \mathcal N_{\vec p}/T}/Z_{\vec{p}}$, where
\begin{align}
    Z_{\vec{p}} = \sum_{\mathcal N_{\vec p}=0}^{\infty} e^{-\omega_p \mathcal N_{\vec p}/T} = \frac{1}{1-e^{-\omega_p/T}},
\end{align}
for a bosonic particle. For a fermion, $\mathcal N_{\vec p}$ can only take a value of $0$ or $1$.
The occupation number is given by
\begin{align}
	\overline {\mathcal N}_{\vec p} = \sum_{\mathcal N_{\vec p}} \frac{\mathcal N_{\vec p} \, e^{-\omega_p \mathcal N_{\vec p}/T}}{Z_{\vec{p}}}
	=-T\frac{\partial}{\partial\omega_p}\ln Z_{\vec{p}} = \frac{1}{e^{\omega_p/T}\mp 1},
\end{align}
where the minus (plus) sign corresponds to the bosonic (fermionic) particle.

\addcontentsline{toc}{section}{Bibliography}
\bibliographystyle{JHEP}
\bibliography{ref}

\end{document}